\documentclass[twoside,11pt]{Latex/Classes/PhDthesisPSnPDF}

\usepackage[T1]{fontenc}
\usepackage{array}
\usepackage{pdfpages}

\usepackage{graphicx}

\usepackage{xcolor}

\usepackage{mdframed}

\begin{document}

\thispagestyle{empty}

\begin{center}

Vrije Universiteit Amsterdam

\vspace{1mm}

\includegraphics[height=28mm]{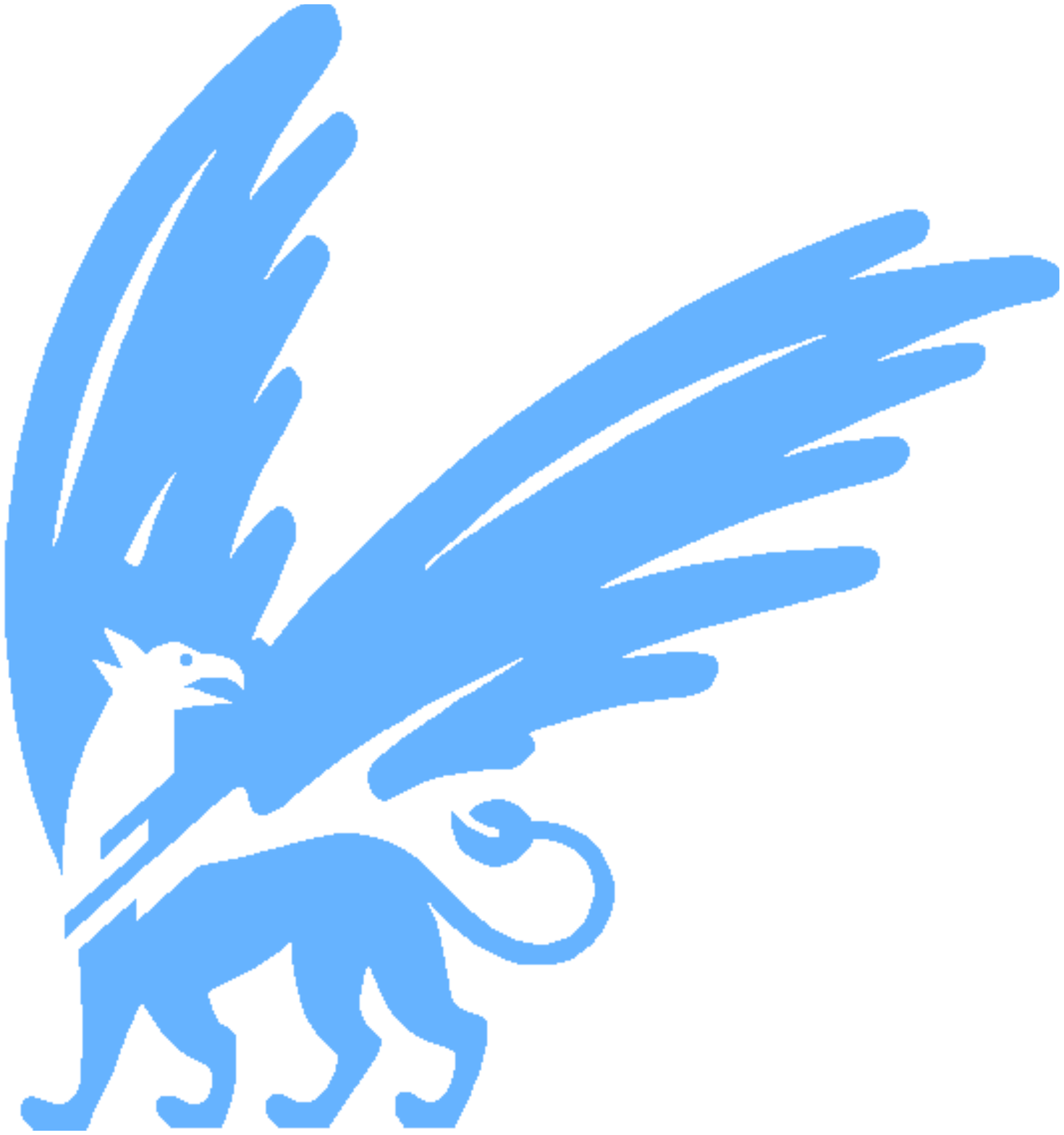}

\vspace{2cm}

{\Large Bachelor Thesis}

\vspace*{1.5cm}

\rule{.9\linewidth}{.6pt}\\[0.4cm]
{\huge \bfseries The OpenDC Microservice Simulator:\par}
{\huge \bfseries Design, Implementation, and Experimentation\par}

\vspace{0.4cm}
\rule{.9\linewidth}{.6pt}\\[1.5cm]

\vspace*{2mm}

{\Large
\begin{tabular}{l}
{\bf Author:} ~~Muhammad Ahsan ~~~~ (2663138)
\end{tabular}
}

\vspace*{1.5cm}

\begin{tabular}{ll}
{\it 1st supervisor:}   & ~~prof. dr. ir. Alexandru Iosup \\
{\it daily supervisor:} & ~~Sacheendra Talluri, MSc \\
{\it 2nd reader:}       & ~~Jesse Donkervliet, MSc
\end{tabular}

\vspace*{2cm}

\textit{A thesis submitted in fulfillment of the requirements for\\ the VU Bachelor
of Science degree in Computer Science }

\vspace*{1cm}

\today\\[4cm] 

\end{center}

\newpage

       
\hbadness=10000
\hfuzz=50pt



\renewcommand\baselinestretch{1.2}
\baselineskip=18pt plus1pt


\begin{center}
\end{center}


\newpage
%
%
%
%





\begin{abstracts}        


Microservices is an architectural style that structures an application as a collection of loosely coupled services, making it easy for developers to build and scale their applications. The microservices architecture approach differs from the traditional monolithic style of treating software development as a single entity. Microservice architecture is becoming more and more adapted. However, microservice systems can be complex due to dependencies between the microservices, resulting in unpredictable performance at a large scale. Simulation is a cheap and fast way to investigate the performance of microservices in more detail. This study aims to build a microservices simulator for evaluating and comparing microservices based applications. The microservices reference architecture is designed. The architecture is used as the basis for a simulator. The simulator implementation uses statistical models to generate the workload. The compelling features added to the simulator include concurrent execution of microservices, configurable request depth, three load-balancing policies and four request execution order policies. This paper contains two experiments to show the simulator usage. The first experiment covers request execution order policies at the microservice instance. The second experiment compares load balancing policies across microservice instances.

\end{abstracts}



\frontmatter


\setcounter{secnumdepth}{3} 
\setcounter{tocdepth}{3}    
\tableofcontents            


\mainmatter

\renewcommand{\chaptername}{} 




\chapter{Introduction}

\ifpdf
    \graphicspath{{1_introduction/figures/PNG/}{1_introduction/figures/PDF/}{1_introduction/figures/}}
\else
    \graphicspath{{1_introduction/figures/EPS/}{1_introduction/figures/}}
\fi



The microservice architecture is widely utilized in entertainment and ecommerce. For example, many well-known companies such as Amazon \cite{amazon-use-ms}, Spotify \cite{spotify-ms}, and Uber \cite{uber-ms} are using microservices. With this style of architecture, these companies claim to have achieved the scalability they need to serve millions of users around the world. Beside scalability, microservices can be both agile and reliable \cite{scalable-ms1, scalable-ms2}. As apps grow larger and more complex, developers need a new development approach that allows them to scale quickly as their needs and demands grow. This is where microservices play their role. Each microservice represents a lightweight and independent functionality with well-defined boundaries. This paper builds a microservice simulator to analyze the performance of microservices. 

\section{Context} 

The monolith architecture is used in traditional applications. It has tightly coupled modules, deployed as a single package and executed as a single process \cite{gamma_helm_johnson_vlissides_1994, DBLP:journals/cloudcomp/TaibiLP17}. The term monolith stands for systems with large volumes of code that run as one process. The monolith architecture has several limitations for large-scale distributed systems, such as limited scalability, dependencies between modules, etc \cite{wolff2016microservices}. 

The consequence of monolith architecture is that every member of the software development team needs to coordinate their activities carefully so they do not block each other because of a single code base. All developers will be blocked if one developer interrupts the build. Furthermore, the code will naturally tend towards deep complex dependencies. The monolith architecture has all-or-nothing deployments. If an older version of the monolith application is running in production and a newer version is staged, a lot of additional work must be done to upgrade the production without impacting the existing business logic.

To overcome the limitations of the monolith, cloud and decentralized applications have moved significantly from monolithic applications to hundreds of loosely coupled microservices graphs \cite{DBLP:journals/cloudcomp/TaibiLP17}. The microservices architecture provides a powerful configuration mechanism. It allows adding ability, where you can add functionality by introducing new parts instead of changing old parts. Microservices are modular, autonomous logical functional units that are independently deployed and scalable. They are transforming the way we work. All agile software development techniques, such as moving applications to the cloud and using containers, are being used with microservices to revolutionize application development and delivery.

Microservices are appealing because they are highly granular, independently built and deployed. As microservice can be independently deployed, they can be independently scaled to respond to varying workloads and user demands \cite{microservice-define-1}. Microservices architecture eliminates a single point of failure through distribution of functionality to various microservices. Even if a single service goes down, the application will continue to work. Additionally, microservices are designed to be loosely coupled with minimal dependency on other services and libraries. As applications tend to become complex, demanding on-the-fly scalability and high responsiveness. Microservices play a crucial role in fulfilling these demands. Moving to a microservices-based approach makes app development faster and easier to manage, requiring fewer people to implement newer features.

\section{Problem Statement}

\begin{figure}[]
\centering
\begin{minipage}{.5\textwidth}
    \centering
    \includegraphics[width=\textwidth,keepaspectratio]{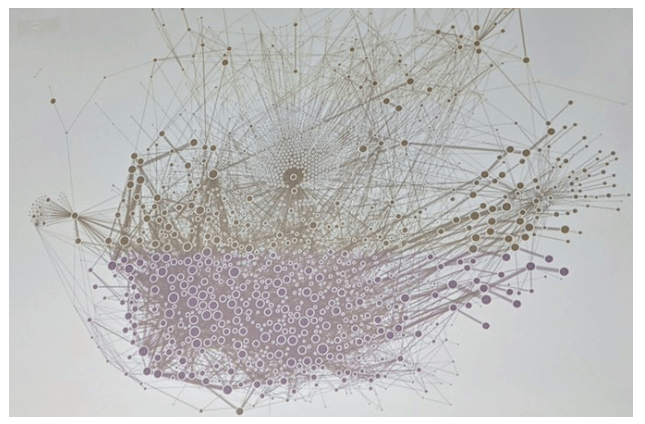}
    \caption{Uber microservice graph \cite{uber-graph}.}
    \label{fig:uber-graph}
\end{minipage}%
\begin{minipage}{.5\textwidth}
    \centering
    \includegraphics[width=0.8\textwidth,keepaspectratio]{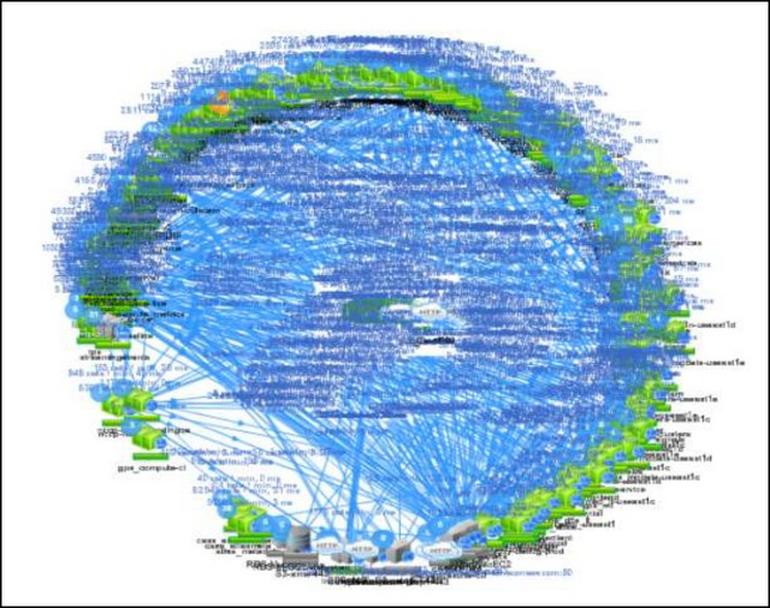}
    \caption{Netflix microservice graph \cite{netflix-graph}.}
    \label{fig:netflix-graphs}
\end{minipage}
\end{figure}

Microservices are continuously more used as developers work to create increasingly \textit{larger}, more \textit{complex} applications that are better developed and managed as a combination of smaller services. The advantages of microservices depend on the quality of building microservices. Figure \ref{fig:uber-graph} and Figure \ref{fig:netflix-graphs} shows examples of microservice graphs for big companies. It can be observed that dependencies among microservices are extraordinarily complex. Due to this, predicting the performance of microservice-based applications is a challenge.

Interactions between microservices are complex. As a result, evaluating and predicting the performance of microservices such as studying the impact of policies and correlating workload models with achievable performance is not straightforward. Although studying these interactions on real test beds provides the most accurate results, several researchers and companies argue that computer \textit{simulation} can be a powerful method to test multiple scenarios and evaluate various policies before enforcing them at different levels \cite{DBLP:conf/globecom/KliazovichBAK10, simulation_benefit_2}. Simulation allows researchers to investigate ``what-if'' scenarios without having to implement them in a data center. In contrast, real-world experimentation is time-consuming, expensive, and has a significant environmental impact. Simulation enables exploration, analysis, and comparison of systems such as simulating schedulers, load balancers, etc. Microservices simulator can provide insight into which variables are most important to the performance of microservices. Simulation can also be used to capture the impact of dependencies between microservices. 

Queuing delay affects the execution time of requests. This occurs when requests are waiting for previously queued requests to finish. The algorithm used to select the request to be executed is called the \textit{request execution order policy}. The duration of the request will change if the requests are executed in a different sequence. Thus, it is important to analyze the effect of request execution order policies on the performance of microservices. 

Load balancer is used to determine the instance of the microservice to route the request to. This algorithm is called as the \textit{load balancing policy}. A load balancer makes sure no one instance is overloaded. Generally, load balancing increases the responsiveness of an application by distributing the work \cite{load-blancing-benefit-2}. As load balancer is an important component of microservice systems, we are interested in comparing microservice performance among different load balancing policies.

\section{Research Questions}

\noindent \textbf{[RQ1] What is a good microservices architecture for the simulator? How does it compare to architectures designed in prior studies?}

A microservice architecture is a set of rules and methods that describe the functionality, organization, and implementation of the microservices. Firstly, a reference architecture must be designed that describes the components of microservice architecture and their roles. The requirements for microservices reference architecture must be defined. The requirements are put together to work out the components of the architecture. This is the most crucial part because components such as authentication and containerization are beyond the scope of the study. The result should be an architecture design that presents the components of the microservice architecture.  \newline

\noindent \textbf{[RQ2] How to implement the components of architecture from RQ1 in the simulator in OpenDC?}

The simulator will follow the microservice architecture from RQ1. The microservice architecture is abstract and thus cannot be directly implemented. Therefore, we need to design a simulator. Firstly, the requirements of the simulator design should be defined. The simulator design should be implementable in OpenDC. Furthermore, multiple methods exist to model a component of the simulator. For example, the load balancing can be client sided or server sided. The design choices and implementation methods for the simulator implementation must be discussed. Following these methods and the simulator design the simulator will be implemented. \newline

\noindent \textbf{[RQ3] What is the impact of request reordering and load balancing policies on the performance of microservice applications?}

The aim of the simulator is that it should provide insights on microservices performance. After building a basic simulator, more advanced features such as request reordering can be added to highlight the usefulness of the simulator. This also makes the output more valuable as it would allow comparison of policies. The implementation from RQ2 is built on to support the advanced features.\newline

\noindent \textbf{[RQ4] How to validate the simulator to show the results are trustworthy? }

The simulator is reliable if the results of the simulator match the actual output. The simulator should be validated to show that its output is reliable. Some validation methods may not be applicable, for example, due to the simulator implementation limitations. Therefore, methods of validation should be discussed.

\section{Approach and Contributions}

\textbf{RQ1.} To answer the first research question, initially the requirements of the microservice architecture are defined. The components of the microservice architecture are worked out. A first approach to decide the simulator components is using insights from architectures from prior studies, then selecting common components and filtering out uncommon internal details \cite{modern-architecture, continuous-software-engineering, Tool-to-Evaluate-Microservice-Software-Architectures}. The idea is to select the fundamental components of a basic simulator then add further components later based on features used in the experiments. The selected components are put into a reference architecture diagram. This diagram will be used as a reference for the simulator design and implementation.

\noindent \textbf{RQ2.} To answer the second research question, firstly the requirements of the simulator design are analyzed. The simulator design and implementation follow the reference architecture from RQ1. The component details for implementing the reference architecture in OpenDC are discussed such as inner parts of the components of the architecture. Multiple ways exist to design and implement the component of the architecture. For example, the load balancer selects an instance of the microservice to run the request on. The load balancer can be implemented on the client or on the server. The choices for the simulator implementation are described in Section \ref{simulator-implementation}. To answer this research question we describe the various methods for implementing the component of the simulator and discuss the pros and cons of each method along with our choice of a method. The methods discussed are mentioned in literature \cite{microservices-design-to-deployment}.

\noindent \textbf{RQ3.} To answer this research question, advance features are included in the simulator for comparison of microservices systems. For example, three load-balancing policies are implemented and evaluated. Three major features are implemented in the simulator. A client request can have multiple paths of microservices calls for that request, see Figure \ref{fig:request-examples}. The depth is the number of microservices visited in the paths. The first feature is that the request can have configurable depth and at any point microservices can communicate with any number of microservices. This feature improves the simulator as different topology can be analysed. The second feature is that the simulator will allow comparison of load balancing policies. The third feature is simulator implements different requests reordering policies. The simulator allows custom workload with configurable interarrival time, execution time, etc. However, the three features mentioned are the highlights of the simulator. Based on these features, two experiments are included in the simulator. The first experiment compares load balancing policies across microservice instances. The second experiment will compare requests execution order policies at microservice instance. In both experiments use the first feature of depth, specifically requests with different depths are used.

\noindent \textbf{RQ4.} The last research question is answered by discussing the methods of validating the simulator. One way is to compare the simulator with another simulator. However no validated microservice simulator is known at this point. The second method is to run a real microservice application, then compare the results with the simulation run. A third approach to validation will be validation using mathematical models. In this study, we validate using the third approach. The simulator output is matched with mathematically calculated result. This is minimal validation and has limitations such as only results that have a mathematical formula can be verified. Details are discussed in Section \ref{limitations} 

The study results in the following contributions:

\begin{enumerate}

\item[\textbf{[C1]}] (Conceptual, Technical) A reference architecture for microservices. The components for the microservice architecture are analysed and put together in RQ1. This resulting into the microservice architecture shown in Figure \ref{fig:ref-architecture} in Section \ref{requirements}.

\item[\textbf{[C2]}] (Technical) A microservice simulator. Based on the reference architecture from RQ1, RQ2 discusses the details of the simulator such as the methods to be used in the simulator implementation in OpenDC. Therefore, at the end of RQ2 the microservice simulator is ready in Section \ref{simulator-implementation}. The simulator uses models and distributions to generate workload.

\item[\textbf{[C3]}] (Experimental) Experiments using the simulator. In RQ3, we use the simulator to perform the two experiments described in Chapter \ref{chapter:evaluation}. The first experiment concerns request ordering at the microservice instance. In the second experiment, load balancing policies are compared.

\end{enumerate}

In this thesis, we design and implement a microservice simulator to understand performance of microservices. We manage to capture in the list below a number of research, education, and business related use cases for our system:

\begin{enumerate}
    \item Users can use this tool to reproduce and evaluate research studies conducted in real world microservice applications. 
    
    \item Users can use this tool to compare load balancing policies, as shown in Section \ref{section:experiment-2}.

    \item Users can compare and evaluate different request execution order policies. 
    
    \item Users can evaluate how different number of instances run under the same or different workload.
    
    \item Users can evaluate microservices and optimize the setup for actual deployment.
    
    \item Users can analyse dependencies between microservices.

\end{enumerate}

The users include microservice researchers. As real microservice deployment have limitations such as cost and time, simulation can be used to research microservice systems. Companies have to ensure quality of service. The companies using microservice architecture can use the simulator to evaluate performance of their applications before real deployment.

\section{Plagiarism Declaration}

I confirm that this thesis work is my own work, is not copied from any other source (person,
Internet, or machine), and has not been submitted elsewhere for assessment.

\section{Thesis Structure}

The first half of this thesis encompasses the conceptual design contributions. Chapter 2 explains the design to the implementation of the simulator. Firstly, the architectural requirements to design a microservice architecture are analysed. Then a more detailed design of the simulator is built. The simulator design provides an overview of the simulator. The design decisions are also discussed. In the second half of this thesis, we present the practical contributions. In Chapter 3 the experiment setup is explained followed by the requests execution order experiment and load balancer experiment. Moreover, we present a summary of the results of the experiments. The limitations and related work are discussed in Chapter 4. Lastly, we end this work summarizing the answers of the research questions and conclusion in Chapter 5. The conclusion summarizes the paper’s work and main contributions. In the end future work options are discussed.



{\let\cleardoublepage\relax \chapter{Simulator Design}


\ifpdf
    \graphicspath{{3_design/figures/PNG/}{3_design/figures/PDF/}{3_design/figures/}}
\else
    \graphicspath{{3_design/figures/EPS/}{3_design/figures/}}
\fi

%

The performance of microservices can be studied by deploying real microservice applications or through simulation. Deploying real microservices to experiment is costly and requires a lot of time, especially for longer experiments \cite{microservice-cost}. These limitations can be overcome by using simulation. By using simulation, experimenting microservice performance can be done without any cost and takes a few seconds to minutes depending on the setup. Theses points clarify the requirement and importance of a microservices simulator. 

OpenDC is a platform for data center simulation \cite{opendc}. It helps understand how data centres work. This study adds microservice simulation to OpenDC. The microservice simulator will be build on OpenDC. For example, microservice instances run on machines that are already implemented in OpenDC. The current project aims to build a microservice simulator to study the performance of microservices and the dependencies between them. To build the simulator, a microservice reference architecture is first designed. This section describes the requirement of the microservice architecture, simulator design, design decisions and implementation of the design in OpenDC. 

\section{Requirements} \label{requirements}

This section answers the first research question that is about the requirements of the microservice architecture. The microservice simulator will be based on a microservice architecture. The microservice architectures designed by prior studies do not meet the requirements of the project. Firstly, their designs are tailor-made for their projects; specifically they contain components that are beyond the scope of the current study. For example, the reference architecture in this study has a similar outline to the architecture in the paper Continuous software engineering – A microservices architecture perspective \cite{continuous-software-engineering}. However, in the current study, the microservice architecture is simplified. For example, we consider all types of microservices as one unlike the paper where they divide into categories. Similarly another architecture by Microservices Architecture For Modern Digital Platforms \cite{modern-architecture} is well structured but contains components such as containerization, caching, etc. Such components are beyond the current paper scope because they are not the fundamentals for the microservice simulator. The reference architecture of current study is aimed at being simple while includes essential components. A stating approach is using insights from prior studies, using common components and filtering out uncommon internal details \cite{modern-architecture, continuous-software-engineering, Tool-to-Evaluate-Microservice-Software-Architectures}. The idea is to select the fundamental components of a basic simulator. Then add further components later based on features used in the experiments. This microservice architecture acts as a guide for the simulator implementation. 

Requirements for microservice reference architecture (where requirement x under this category is labelled A.x). The architectural requirements are based on insights from previously mentioned literatures about microservice architecture. 

\begin{enumerate}

\item[A.1] The design should have the fundamental components which are described as the component common in prior studies \cite{ microservices-design-to-deployment, modern-architecture}.

\item[A.2] The design should support routing to handle requests. 

\item[A.3] The design should support load balancing across microservice instance.

\item[A.4] The design should support forwarding request to microservices instances for execution.

\item[A.5] The design should support communication between microservices. 

\end{enumerate}

Simulation design requirements, on the other hand, are mostly composed of customizable platform specifics such as customizable depth of requests (labeled S.x).

\begin{enumerate}
    \item[S.1] The simulator should support easily configurable inputs such as interarrival time, execution time, etc.
    
    \item[S.2] The system should support tracking metrics from the simulation.
    
    \item[S.3] The simulator should support concurrent execution of microservices.
    
    \item[S.4] The simulator should support configurable depth of requests.
    
    \item[S.5] The simulator should support multiple load balancing policies.
    
    \item[S.6] The simulator should support multiple request execution order policies.
    
    \item[S.7] The system should support repeatable executions and controlled randomization through the use of deterministic seeding methods.
    
    \item[S.8] The system should compile a report of the recorded metrics at the end of the simulation.

    \item[S.9] The system should support running simulation with and without workload trace.
\end{enumerate}

Regarding the requirement of running the simulator using trace, there is no trace currently available so the simulator implementation uses statistical models for workload generation. The simulator implementation can be adapted such that the statistical models are replaced so trace can be used. Table \ref{tab:ms-trace-format} shows the expected minimal trace format. The format also helps to understand the current simulator implementation as the current workload generation have similar inputs. Each row represents a microservice request. The request Id is unique id for the client request. Microservice requests that are part of a client request have the same `request id'; thus multiple rows can have the same request id. Timestamp is the time of the microservice invocation request. The microservice called is in the `called ms' field supported by the `called by' field which indicates the microservice that called the current microservice. For depth 0 the called by field is null as those are the starting microservices. Hops done field is the depth of the current request. The field `exetime' is the execution time of the current microservice in microseconds. The trace is minimal, other fields such as CPU usage can be added.

\begin{table}[]
\centering
\resizebox{0.9\textwidth}{!}{%
\begin{tabular}{|l|l|l|l|l|l|}
\hline
request Id & timestamp & called ms & exetime & hops done & called by \\ \hline
\end{tabular}%
}
\caption{Trace format for the microservice simulator.}
\label{tab:ms-trace-format}
\end{table}

\section{Reference Architecture for Microservice-based Operations}

This section discusses the components for the reference architecture based on the architectural requirements defined in Section \ref{requirements}. The literature Microservices From Design to Deployment by Chris Richardson \cite{microservices-design-to-deployment} provide the base for discussion of different methods along with other studies mentioned in the description of components.Figure \ref{fig:ref-architecture} shows the reference architecture. 

\begin{figure}[ht]
    \centering
    \includegraphics[width=.8\textwidth,keepaspectratio]{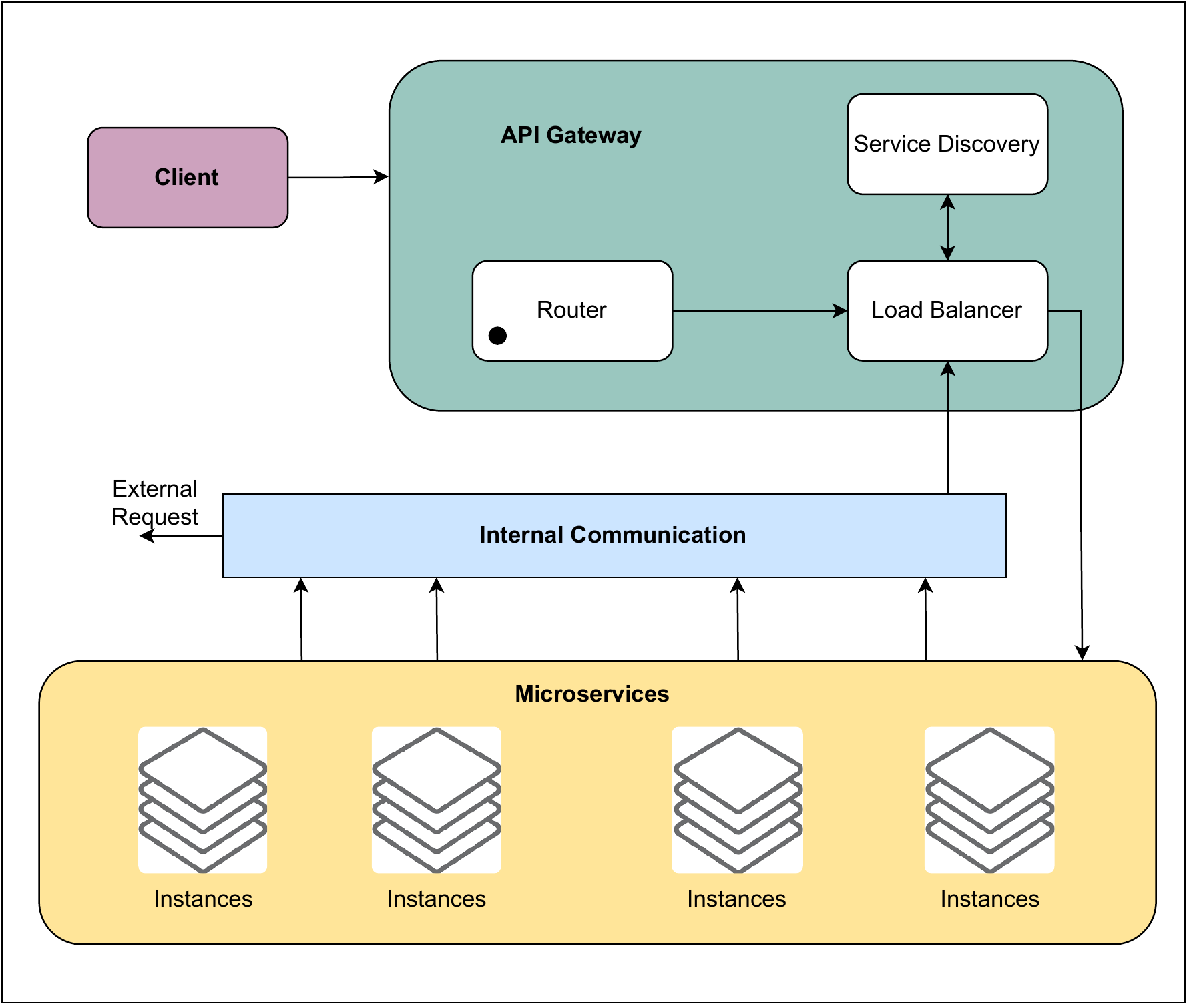}
    \caption{Microservice reference architecture.}
    \label{fig:ref-architecture}
\end{figure}

\textbf{Client} consists of consumers such as mobile apps, web applications, IoT devices, third-party services, etc. The client sends the request to the API Gateway.

\textbf{API Gateway} provides a single endpoint or URL for the client apps and then internally maps the requests to a group of internal microservices \cite{microservices-design-to-deployment}. 

There are two ways for clients to send requests to the microservices \cite{api-gateway-service-discovery}. One way is to send a request directly to the microservices. Such implementation does not have an API Gateway to map requests and require the client to send separate requests to microservices. However this approach has limitations. One problem is that the client has to make separate requests to each microservice and keep proper track of their order. This makes the client code much more complex. Another problem is that, overtime if we need to make changes to how the system is partitioned such as splitting one service into two separate services, for example splitting order service into package tracking service and a payment service. Then we need to make changes to the client as well, using the example above; originally client sends a request to order service but in the updated back end replacing order service with tracking and payment service, the client needs to be updated to use them. Performing this refactoring every time the back end updates can be extremely difficult.

A much better approach is to use API Gateway \cite{api-gateway-1, api-gateway-2}. In this approach the client sends a request to the API Gateway which then routes request(s) to the appropriate microservices. We use this approach in the simulator. The API Gateway encapsulates the internal system architecture and can provide an API that is tailored to each client. The client is independent of any changes at the back end as API Gateway takes care of the changes by updating the request map. For example, to use the newly added services, the API gateway will update the request map while the client can keep using the same request. Moreover, API Gateway might have other responsibilities such as authentication, monitoring, load balancing, caching, etc. In the current microservice simulator implementation, the API Gateway is composed of router, load balancer and service discovery. Each of these internal components is described later.
 
\textbf{Router}: when the request is received, it is the router's job to determine which microservices should be invoked. As a part of API Gateway, a single request can result into multiple microservice requests to the same or different microservices. For example, a shop app sends a request to retrieve information for a product. The API Gateway then maps this request to multiple services such as product description service, image service, review service, recommendation service, etc. Each of the service may depend on other services such as recommendation service may call history service to make the recommendations for the user. The client request completes when all the microservices and their dependent services have responded. The complete response is then sent back to the client. We define the whole process as a single request from the client that the paper calls the client request. Each microservice call in the client request is called the microservice request or microservice stage. The microservice request has attributes such as the depth at which the microservice was invoked.

Figure \ref{fig:request-examples} shows examples of client request execution paths. The simulator can run microservices in sequential and parallel. Example 1 is sequential execution while examples 2 and 3 involves parallel executions. The client request is called the complete request path, including all sub paths. In the diagram, the microservices are represented by the yellow circles. Microservice request or microservice stage is the arrow going to the microservice. The diagram also shows the communication of microservices and their depths.

\begin{figure}[ht]
    \centering
    \includegraphics[width=.7\textwidth,keepaspectratio]{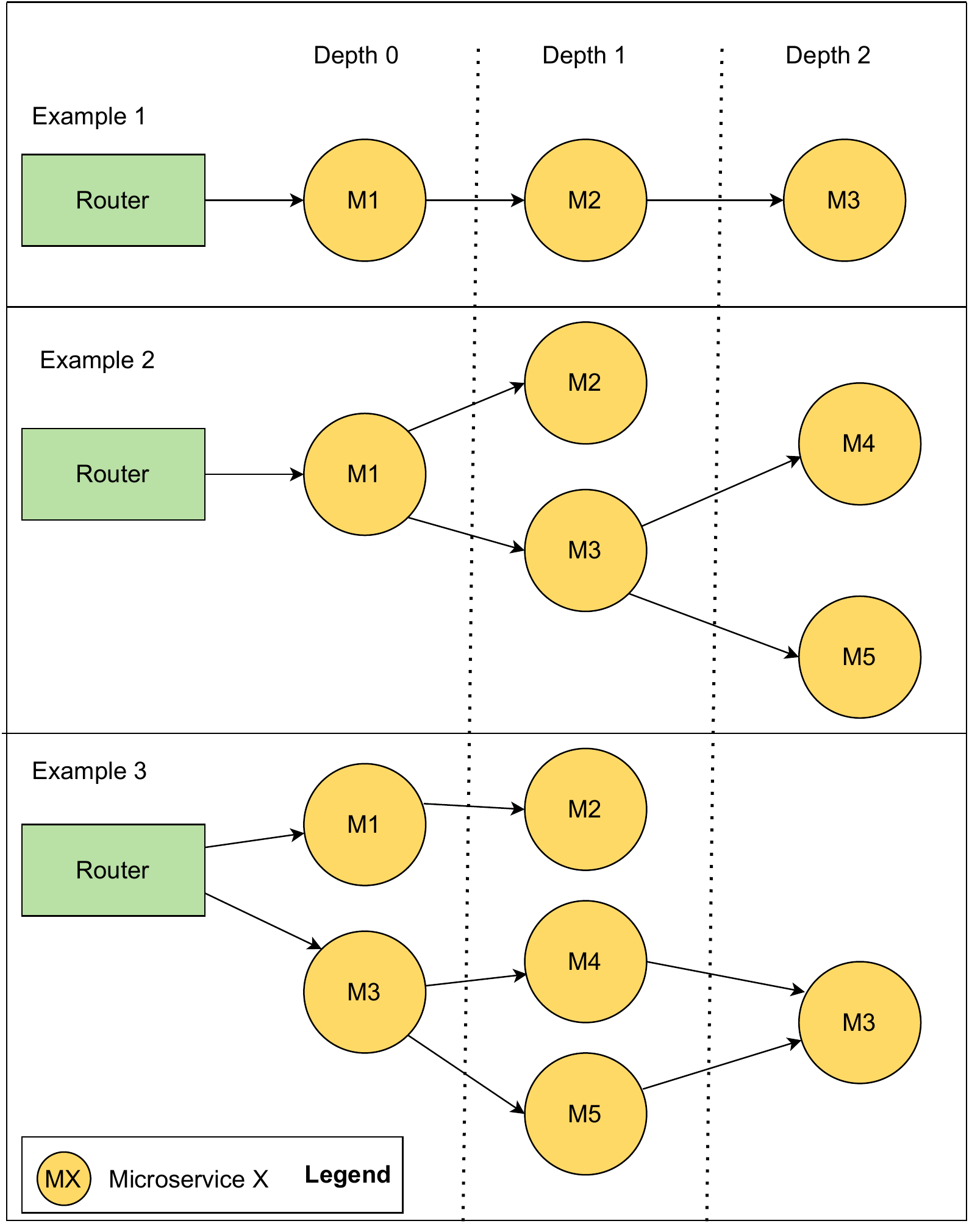}
    \caption{Example of requests to simulator and their paths.}
    \label{fig:request-examples}
\end{figure}

\textbf{Load Balancer} is responsible for distributing the microservice requests to the instances of that microservice, ensuring no instance is overloaded, under-loaded or idle \cite{load-balancing-definition-1, load-balancing-definition-2}. Load balancing tries to speed up different constrained parameters like response time, execution time and system stability \cite{load-blancing-benefit-1}. Furthermore, load balancing can be implemented on the client and on the back end \cite{load-balancing-client-server}. In this study, the simulator simulates load balancer on the back end as part of the API Gateway. Server sided load balancing is preferred because it simplifies the client. This study excludes the scaling of load balancers. It is assumed that no delay is caused by the operations of the API Gateway. A variety of load balancing algorithms exist. Details of load balancing algorithms used in the simulator are mentioned in Section \ref{simulator-implementation}.

\textbf{Service discovery} maps logical service names to physical addresses \cite{service-discovery-1}. The set of running microservice instances changes dynamically. Instances have dynamically assigned network locations. Therefore, service discovery tracks which instances are active and their physical locations in the service registry. The load balancer consults service discovery to route the request to the microservice instance. 

If load balancing is client sided, then service discovery can be assessed directly by the client. The client retries the active instance information from service discovery to perform load balancing on the client. This method of service discovery is known as client-side discovery. On the other hand, if load balancing is server sided then client does not know of service discovery, clients make requests via a router, the load balancer queries the service discovery and forwards the request to an available instance. This method of service discovery is known as service-side discover \cite{service-discovery-1}. In this project server sided load balancing with service-side discovery is used. Therefore both service discovery and load balancer are part of API Gateway as shown in Figure \ref{fig:simulator design}.

\textbf{Microservices}: In microservice architecture, the application is divided into separate services. Each service of the application operates as an independent service \cite{tao-of-microservice}. For each microservices there can be different number instances running depending on the load. The instance on which the request runs is decided by the load balancing algorithm. The number of active microservice instances can be dynamic as they can be scaled depending on the load. The location of the instances can also change. Instances are registered, updated and deregistered at the service discovery. There are two main ways to keep the service discovery in synchronization with the active instances. One method is that instance sends updates to the service discovery of their status. The other method is the service discovery checks periodically instance status. Any method can be used depending on the preference, for example if preference is to keep instances simple then second method can be used.

\textbf{Communication}: Microservices may need to communicate with other microservices. They may also need to communicate with external APIs. Furthermore the communication may be synchronous or asynchronous. Communication may include single or multiple requests.

In this project the communication is implemented such that after the current microservice executes, requests are forward to (invoke) other microservices. The current microservice is done and can continue execution of other requests. To simulate the response back to the current microservices, the forwarded microservice can later call the current microservice again after their execution. This setup of communication uses asynchronous communication approach. This approach is more flexible and does not have to worry about being stuck in a cycle. This allows studying client requests with high depths.

\newpage

\section{Design For OpenDC} \label{simulator-implementation}

In the current and the next section, we answer the second research question. This section describes the design of the simulator in OpenDC. The design is based on the reference architecture and the simulator design requirements from Section \ref{requirements}. There is no known microservice trace at the time of the current study that can be used as input for the simulator. Therefore the implementation of the simulator uses models and distribution to run the simulator without the need of trace. The simulator provides a simple interface to replace the existing customizeable models and policies which are indicated by dashed boxes in Figure \ref{fig:simulator design}. The figure shows a high-level overview of the simulator.

\begin{figure}[]
    \centering
    \includegraphics[width=\textwidth,keepaspectratio]{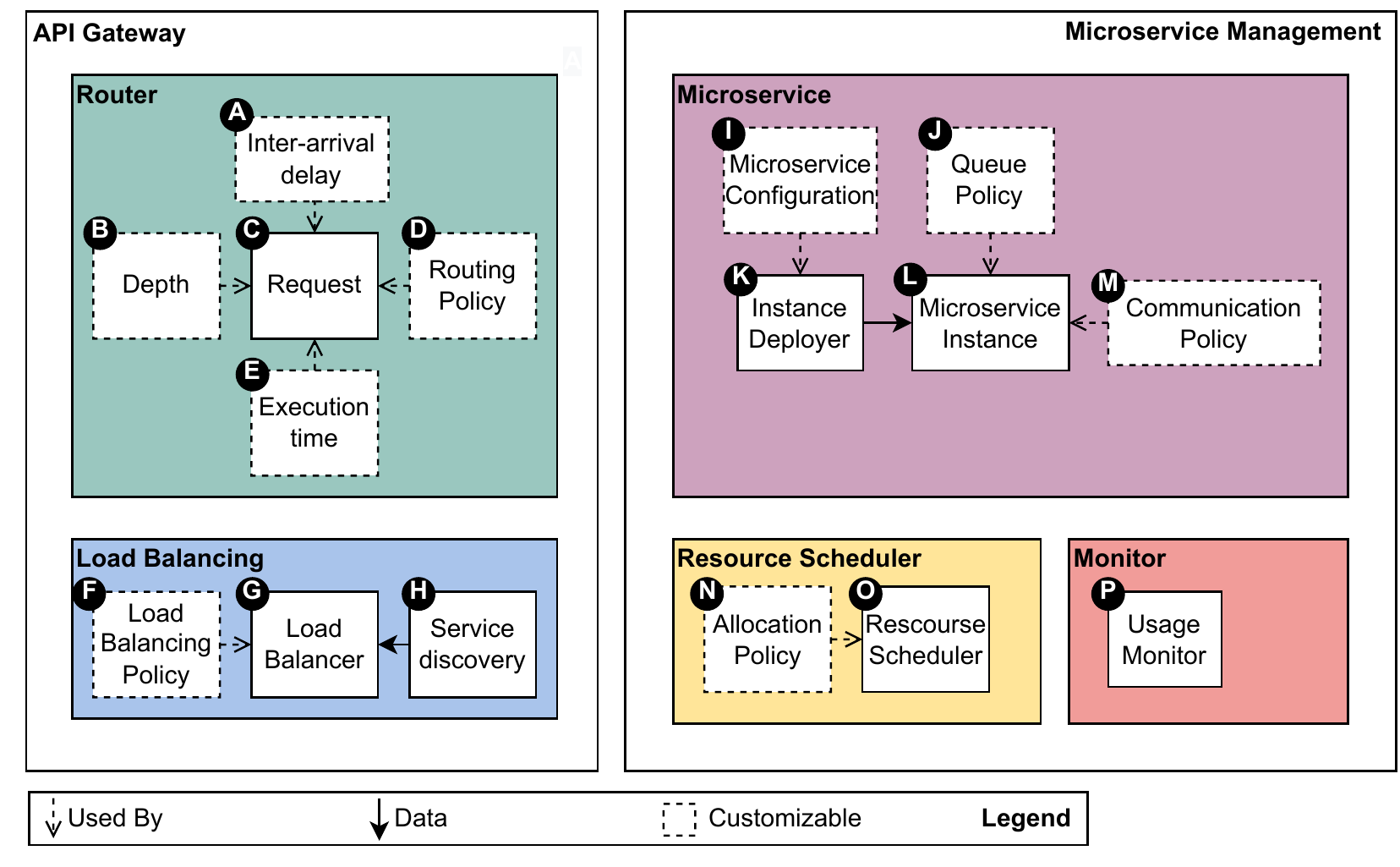}
    \caption{Microservice simulator overview.}
    \label{fig:simulator design}
\end{figure}

Client request is built by the router using the simulator configurations. The configuration for the router includes inter-arrival delay, depth, execution time and the routing policy. The inter-arrival delay (\textbf{A}) is the delay between client requests. Poisson arrival is the default implementation used for inter-arrival delay \cite{poisson-arrival}. It simulates the arrival of client requests based on the rate value passed to the Poisson distribution. The depth (\textbf{B}) is the maximum depth of the client request. If microservices do not call other microservices the depth is zero. By default depth of request is sampled based on probability such as probability is 50\% request have depth 2 and remaining 50\% have depth 0. Request (\textbf{C}) or the client request is the full map of the path of the request. It consists of all the microservices that would be invoked. Examples of client request are shown in Figure \ref{fig:request-examples}. To simply state, a client request is made of microservice requests.

The routing policy (\textbf{D}) provides the microservices that would be invoked at depth 0. The default routing policy is probability based routing. By default the policy selects microservices using the invoke probability of the microservices given the number of microservices to invoke. Execution time (\textbf{E}) for each microservice invocation is included in the microservice request. By default sample from log-normal distribution is used to fill execution time \cite{slowdownPaper}. The log-normal distribution is an efficient distribution to model the size distributions of various phenomena and has a higher value for the upward distribution than a normal distribution \cite{lognormal-1, lognormal-2}.

The load balancer (\textbf{G}) uses an algorithm to decide on the instance of microservice to send the request to. Server sided load balancing is used in the simulator. The simulator implements three load balancing policies (\textbf{F}) which are greedy load balancing \cite{greedy-1}, least connection load balancing and round robin load balancing algorithm \cite{load-balancing-types, load-balancing-client-server}. The greedy load balancer chooses the instance of microservice that has the least load. The load is the total execution time of queued requests and the remaining execution time of the currently executing request at the instance. Therefore to use the greedy load balancer, execution time has to be known for example by using an execution time prediction model. The least connection load balancer chooses the microservice instance which has the least number of active connections. Active connections is the number of requests queued at the instance. Lastly, the round robin load balancer goes down the list of microservice instances in the group. The round‑robin load balancer forwards a client's request to each instance in turn. When it reaches the end of the list, the load balancer loops back and goes down the list again. The Service discovery (\textbf{H}) stores the active microservice instances. The load balancer uses service discovery to fetch the active microservice instances to select the instance. Furthermore, microservice instances update the service discovery when they are deployed and removed.

Microservice Configuration (\textbf{I}) is the number of microservices and amount instances for those microservices. This configuration is passed to microservice instance deployer (\textbf{K}) that would deploy the instances. The current simulator deploys the instances before the simulation starts as scaling is not supported. After deployment the microservice instance (\textbf{L}) starts and waits for requests. The execution order of requests at microservice instance is decided by the queue policy (\textbf{J}). There are four policies implemented in the simulator. First is first come first serve (FCFS) in which requests the arrives first is executed first \cite{first-come-first-serve}. Second policy is non-premitive shortest job first (SF) which gives priority to requests with smaller execution time. Third policy is fair share of resources (FS) in which request is executed for a specific time and if not completed, the request is put back into the queue. The fourth policy is early deadline policy (ED) where deadline is the deadline for the microservice request. This policy has two variants depending on the method of setting the deadline. The deadline for the first variant is set by equally dividing the slack between microservice stages called equal division of slack (EDS). 

The second variant divides the slack according to execution time, then set the deadlines using the slack values, we call this execution based division of slack (EXDS). These variants are inspired by the GrandSLAm paper \cite{GrandSLAm}. GrandSLAm paper also uses slack based division. However it focuses on batching which is beyond the scope of current study. In this paper we study the slack-based division policy without batching. The First come first serve, shortest job first and equal division of slack do not require execution time but other policies need to know execution time. In production, the execution time may be obtained from prediction models. After execution finishes request is forwarded to the next microservice, if any. The forwarded microservices are selected using the communication policy (\textbf{M}). The communication policy implementation is similar to router policy. By default, it also uses probability. The probability is the probability of microservice to be called by another microservice. It is reasonable to assume that microservice cannot call itself. However it may later be called by another microservice.

Microservice instances run on machines already implemented in OpenDC which are provided by the Recourse Scheduler (\textbf{O}). The allocation policy (\textbf{N}) is customizable so machine resources can be configured for example the number of CPU, memory of the machine, etc. 

The last component in the Figure \ref{fig:simulator design} is the usage monitor (\textbf{P}). The simulator's metrics are tracked via the usage monitor. The request listeners such as router and microservice instances record metrics include total time, wait time, slow down, and execution times. The metrics recorded by the microservice instances are for microservice requests while the router recorded metrics are for client requests. The metrics recorded by the instances exceed those of the router since communication requests are not included in router metrics. Each microservice hourly utilization and final mean utilization is recorded. These measurements can be used to do experiments and assess the outcomes, such as comparing different policies.

\section{Implementing the design in OpenDC}

\begin{table}[]
\centering
\resizebox{\textwidth}{!}{%
\begin{tabular}{l|l|l} \hline
\textbf{Inputs}                                                                & \textbf{Description}                                  & \textbf{Default}                                                                                    \\ \hline 
End time                                                              & Time limit for simulation                    & 24 hr                                                                                       \\
Execution time                                                        & Execution time of microservice invocation    & Log normal distribution                                                                     \\
Inter-arrival delay                                                   & Delay between client requests                & Poisson Arrival                                                                             \\
Depth                                                                 & Maximum depth of client request.             & Probability for depth                                                                                      \\
SLA                                                                   & Total deadline for request completion        & 4 second                                                                                    \\
Call Microservices                                                    & Microservies that client calls           & \begin{tabular}[c]{@{}l@{}}Probability for microservices \\ to be called\end{tabular}       \\
\begin{tabular}[c]{@{}l@{}}Communication-\\ Microservices\end{tabular} & Microservice to forward the request to & \begin{tabular}[c]{@{}l@{}}Probability for microservices \\ to be communicated\end{tabular} \\
Load Balancer                                                         & Load balancing alogorithm                    & Round Robin                                                                                 \\
Queue Policy                                                          & Algorithm for queued request execution order &   First come first serve                                                                        \\ \hline                
\end{tabular}%
}
\caption{Simulator inputs.}
\label{table:simulator-inputs}
\end{table}

This section implements the design details mentioned in Section \ref{simulator-implementation}, resulting in OpenDC Microservice simulator. The implementation of the simulator is described. To begin with, the simulator parameters must be set. These parameters include  load balancer, maximum depth, arrival delay, execution time, request SLA, microservices configuration; the number of microservices, the number of instances of each microservice, policy for executing the request and the end time. The simulator inputs are shown in Table \ref{table:simulator-inputs}.

After the inputs are provided and before the simulation starts, the microservices are built and their instances deployed. Furthermore the router starts listening to requests. Then simulation starts, initially a client request is built. Microservice request for each stage with attributes such as execution time is included in the client request. The zero-depth microservice caller is null. The maximum depth parameter cannot be exceeded by any of the request path. After the request is made, the simulator will execute it. Starting from 0 depth, executing microservice requests, checking for other microservices that may need to be invoked and so on.

When invoking a microservice, the router uses a load balancing algorithm  to select the appropriate microservice instance. Then the selected instance is invoked. The instance adds the request to its queue. At the instance the requests in queue are executed according to the queuing policy. When execution is complete, the instance checks the client request. If more microservices need to be called, a microserivce request is sent to the router to invoke the microservices. The other case is when no microservices would need to be called. In either case, the current instance is available and can further handle other requests that are waiting in the queue. In parallel, after the inter-arrival interval passes, the simulator builds a new request. These operations will continue until  the simulation runtime is reached or no more requests. At the end of the simulation, the simulator outputs the recorded metrics.

\section{Analysis of How the Design Meets the Requirements}

This section analyses how the simulator implementation meets the design requirements mentioned in Section \ref{requirements}:

\begin{enumerate}
    \item[S.1] The simulator includes configurable parameters such as load balancer, maximum depth,
arrival delay, execution time, request SLA, microservices configurations; the number of
microservices and the number of instances for each microservice, policy for executing the request
and the end time. The simulator inputs are shown in Table~\ref{table:simulator-inputs}. These parameters can be easily configured at the start of the simulation.
    
    \item[S.2] The simulator records metrics for every request and at regular intervals such as hourly utilization of microservices.
    
    \item[S.3] A microservice can communicate with multiple microservices which execute concurrently.
    
    \item[S.4] The depth of the requests can vary during the simulation. For example, the experiment in Section \ref{section:experiment-1} has requests with depths 0 and 2.
    
    \item[S.5] The simulator supports three multiple load balancing policies; round robin load balancer, greedy algorithm and least connection load balancer.
    
    \item[S.6] The simulator supports four request execution order policies, described in Section~\ref{section:experiment-1}.
    
    \item[S.7] The system supports setting a seed for statistical models that use random generation, for example, to sample from a distribution. This allows repeatable execution.
    
    \item[S.8] At
the end of the simulation, the simulator outputs a report. The report contains results
such as microservice utilization, total time, wait time, execution time, slowdown, etc.
The results can be compared to analyse microservices performance.

    \item[S.9] The system does not meet this requirement because currently the simulator only runs using statistical models. Running the simulator using workload traces is not implemented. However the simulator can be adapted to use workload traces, see Section \ref{limitations}. 
    
\end{enumerate}

}




{\let\cleardoublepage\relax \chapter{Evaluation} \label{chapter:evaluation}


\ifpdf
    \graphicspath{{5_evaluation/figures/PNG/}{5_evaluation/figures/PDF/}{5_evaluation/figures/}}
\else
    \graphicspath{{5_evaluation/figures/EPS/}{5_evaluation/figures/}}
\fi

%

This chapter answers the third research question about the impact of request reordering and load balancing policies on microservices performance. Two experiments are run to demonstrate the simulator's operations. The first one concerns request execution order at the microservice instance. In the second experiment, load balancing policies are compared.

\section{Experimental Setup} \label{section:experiment-setup}

\begin{figure}[]
    \centering
    \includegraphics[width=0.7\textwidth,keepaspectratio]{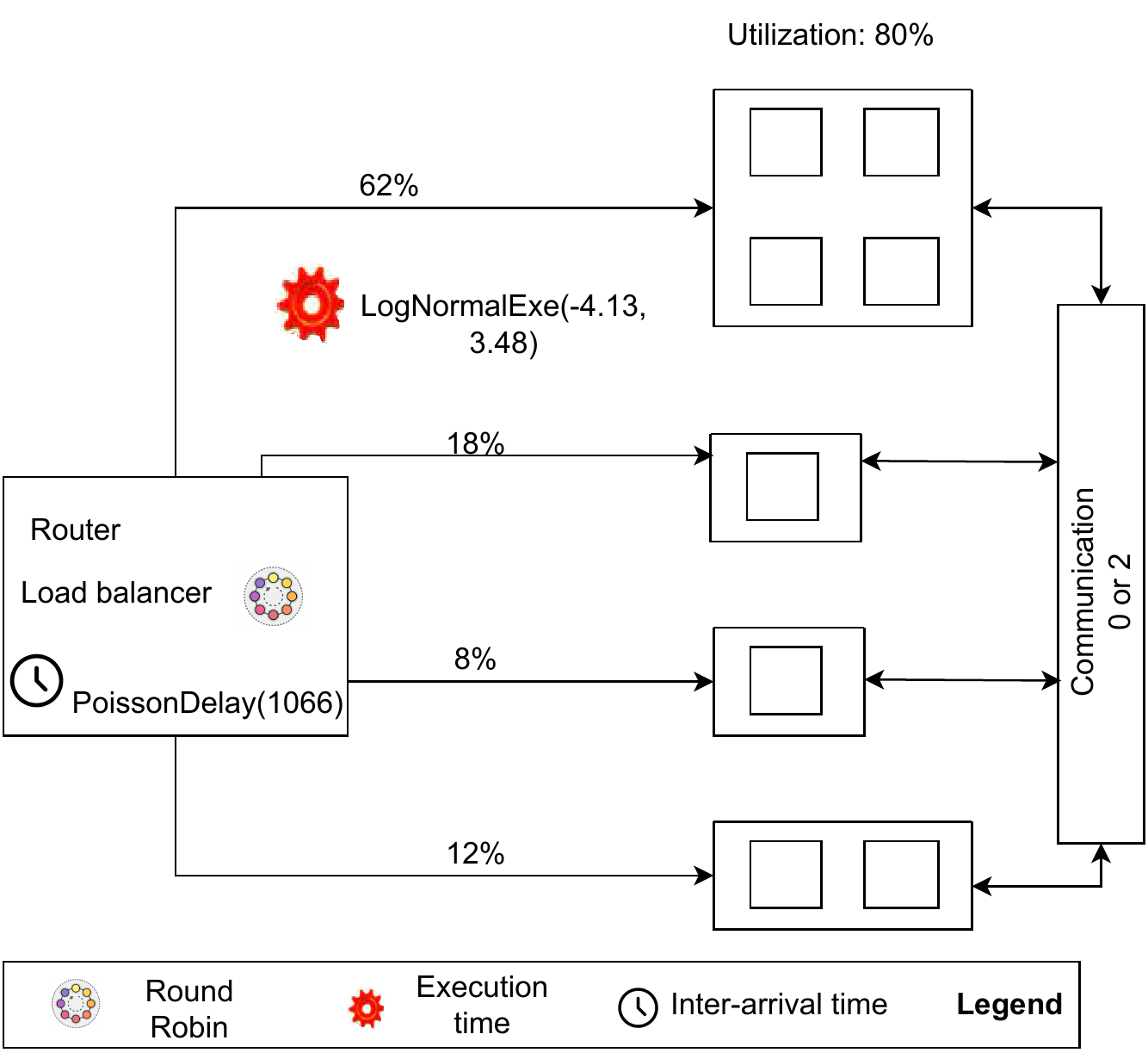}
    \caption{Set up for experiments.}
    \label{fig:experiment setup}
\end{figure}

This section describes the setup for the experiments. The setup mainly includes the input of the simulator. A similar setup applies to both experiments with some differences. The microservice usage is 80\%. It is recorded at hourly intervals and averaged at the end to get the overall utilization for each microservice. The utilization is calculated as the sum of the load at the instances of the microservice during the time divided by the result of time multiplied by the number of instances. The distribution of microservices is; four microservices with a different number of instances depending on the load. One microservice has four instances, another microservice has two instances, and the other microservices each have one instance. For microservices, the call probabilities are 62, 18, 8, and 12 percent. These probabilities are based on the 2021 Azure Trace. Values from the trace are used such that at least 10,000 requests each day, ensuring that microservices aren't utilized very low. Poisson distribution is used to calculate the delay between arrival times with an average of 1066 microseconds. The execution time of microservice request is calculated using a log-normal distribution with mean 4.13 and standard deviation  3.48. These numeric values used to set model and distribution parameters are calculated from 2021 Azure trace \cite{Azurepublicdataset2021}. The simulation time for these tests is 24 hours, so only the trace of day one is used.

The microservices simulator currently does not support scaling. Therefore microservice instances are deployed beforehand. The probability distribution is used to determine the depth of the request. For example, 50\% of requests to have a depth of 2 while rest 50\% have depth zero. A depth of 2 requires the invoke of 3 microservices. The probabilities of depths were chosen as the values to demonstrate the operation of request reordering policies. The result of an equal division of slack policy can only be displayed only if the client requests depth were different causing different stage deadlines. One microservice is called at each depth to control the experiment. Thus, in the experiment the microservice calls are sequential for all depths. Figure \ref{fig:experiment setup} shows the setup.

The difference between the setup of the two experiments is that the first experiment varies request execution order policies, however, in the second, this policy was kept as first come, first serve. Similarly, the second experiment varies load balancing policies while the first experiment uses fixed load balancing policy that is round robin. For simplicity other settings remain the same for both experiments. The metric used to compare the policies in both experiments is slowdown. Total request time divided by actual execution time equals slowdown \cite{slowdownPaper}. Performance is worse when slowdown is higher. The slowdown is used as the metric because it is derived from both wait time and execution time.

\section{Experiment 1: Request Execution Order } \label{section:experiment-1}

Queuing delay affects the execution time of requests. This occurs when requests are waiting for previously queued requests to finish. The execution sequence of requests, the expected execution times of individual requests, and its preceding requests may all be used to quickly determine the queuing delay at runtime. The duration of the request will change if the requests are executed in a different sequence. SLA violations can be decreased, for instance, if shorter requests are much more frequent than bigger ones. In this case, if shorter requests are given priority, more requests will be fulfilled in accordance with the SLA since they do not have to wait for larger ones. Our first experiment is about understanding the slowdown of requests for microservice execution due to execution order. When requests arrive at the instance, they are placed in a queue. The queue policy selects the request to execute. Thus affects the duration of a request in the system.

In the first experiment, we compare four queuing policies, by trying them in the microservice platform, in turn and independently. The trivial \textit{first come first served} (FSFC) policy states that the request that is received first shall be executed first. The second policy is \textit{fair share of resources} (FS). As with first come first served, the fair share policy handles requests in the same order. However, a request can only execute for a given span of time, in this case 500 microseconds. The request will be moved back to the end of the queue if it is not completed. The time of 500 microseconds is set since the distribution of requests execution times reveals majority requests have execution time of 1000 microseconds. A greater number than 500 microseconds in the current setup would reflect less sharing. The third policy is \textit{shortest job first} (SF). It prioritizes requests with shorter execution time. The policies discussed till now are well known. Through this experiment, we would like to evaluate how well they function when used for requests execution order at the microservice instance.

The publication GrandSLAm: Guaranteeing SLAs for Jobs in Microservices Execution Frameworks \cite{GrandSLAm} served as inspiration for the fourth policy, which is more microservice-related. The goal of this policy is to establish deadlines for each microservice stage or slack. The maximum amount of time a request may spend in a specific microservice stage is known as microservice stage slack. The requests are executed in such a way that the request with the earliest deadline is chosen to be executed, this policy is known as the ``early deadline policy'' in this study. In addition, this policy has two different variants depending on how the deadline is determined. These variants also make this policy appealing. 

\begin{figure}[]
    \centering
    \includegraphics[width=.7\textwidth,keepaspectratio]{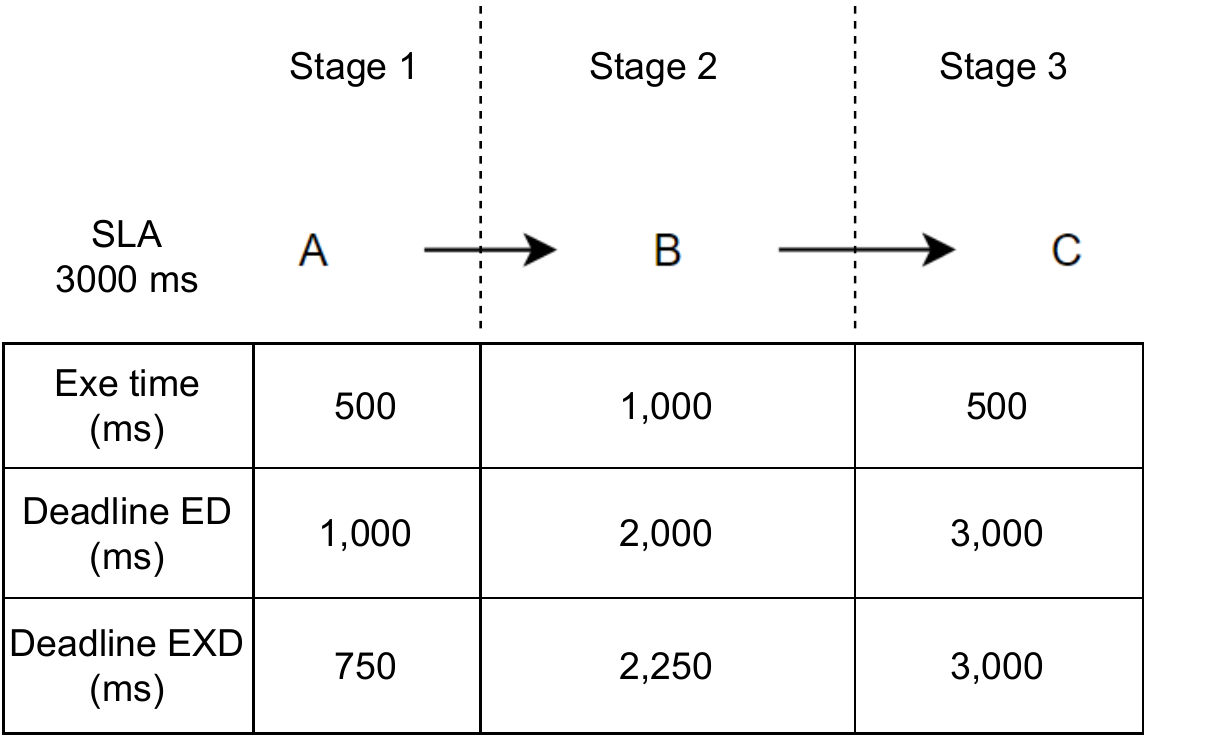}
    \caption{Example of deadlines set using two variants of early deadline policy.}
    \label{fig:early-deadline-variants}
\end{figure}

The idea behind the early deadline policy is that requests that can wait should do so, even if they arrive early. Although the request can be fulfilled early, we wish to use this policy to give priority to those requests that cannot wait or are more urgent. When the earlier request can no longer be delayed—that is, when its deadline gets closer—the request will be executed. Since it makes sense to run urgent requests early, this notion is acceptable. The GrandSLAm paper only employs slack-based policies and investigates sequential execution, whereas the simulator used in this study also permits parallel execution of these policies. However, as in the Grand Slam publication, we exclusively used sequential execution for experimental purpose.

\textit{Equal distribution of slack} (EDS) is the first variant of the equal deadline policy. This policy allocates an equal amount microservice stage slack given the total SLA. The maximum depth of the request is taken into account while allocating slack for the microservice stage. Consider, for example, that two requests, one with depth zero and the other with depth 2. We assume that the microservice instance is currently processing an older request. At the time 6000 microseconds (ms), the request with depth zero is received by the instance. After 500 microseconds, a request with a depth of 2 is received by the instance at time 6500 ms. The slack is shared according to the equal distribution of slack policy. If the SLA is 3000 microseconds, the maximum depth is 2, and the microservice stages are the maximum depth plus one to encompass depth 0. Therefore, the microservice stage slack for each stage is set at 3000/3, or 1000 microseconds. For request with depth 2, the deadlines using the stage slacks are set at 7000 ms, 8000 ms, and 9000 ms for depths 0, 1, and 2, respectively. On the other hand, the slack for the microservice stages at depth 0 is 3000 microseconds which results in a deadline of 9000 ms.

We just focus on the first stage of the two requests in order to simplify the example. As mentioned above, the request with depth 2 has the first deadline at 7000 ms, whereas the deadline for the request with depth 0 is 9000 ms. When the instance is free, it will check on the next request to execute from the queue policy. The depth 0 request arrived earlier. However it can wait because it has a later deadline. Due to the shorter deadline, the depth 2 request is processed first. 

The second variant of the early deadline policy is that the deadline is set using \textit{slack distribution based on execution time} (EXDS). The idea is same as the first variant, that is to give priority to early deadlines, except that in the second variant shorter microservice stages should get shorter deadlines instead of equal division. This can be seen as enhancing the first policy. However the trade-off is that execution time has to be known. The slack division with deadline calculation can be seen in the example shown in Figure \ref{fig:early-deadline-variants} along with comparison to the first variant. The slacks are 750 ms, 1000 ms, 750 ms leading to deadlines 750 ms, 2250 ms, and 3000 ms for stages A, B, and C respectively. 

Since just one microservice is called at each level, the steps in the example are sequential. The policies still function even in the case of parallel executions. In such a scenario, the deadline for the EDS is calculated by dividing using the maximum depth rather than depth of individual inner path. The deadline for the microservice stage is defined using the maximum execution time when using the EXDS variant in parallel setting. While the present simulator also allows parallel execution, the GrandSLAm paper solely employs sequential execution. However, for simplification, the experiment design indicated above uses sequential.

\begin{figure}[]
    \centering    \includegraphics[width=.7\textwidth,keepaspectratio]{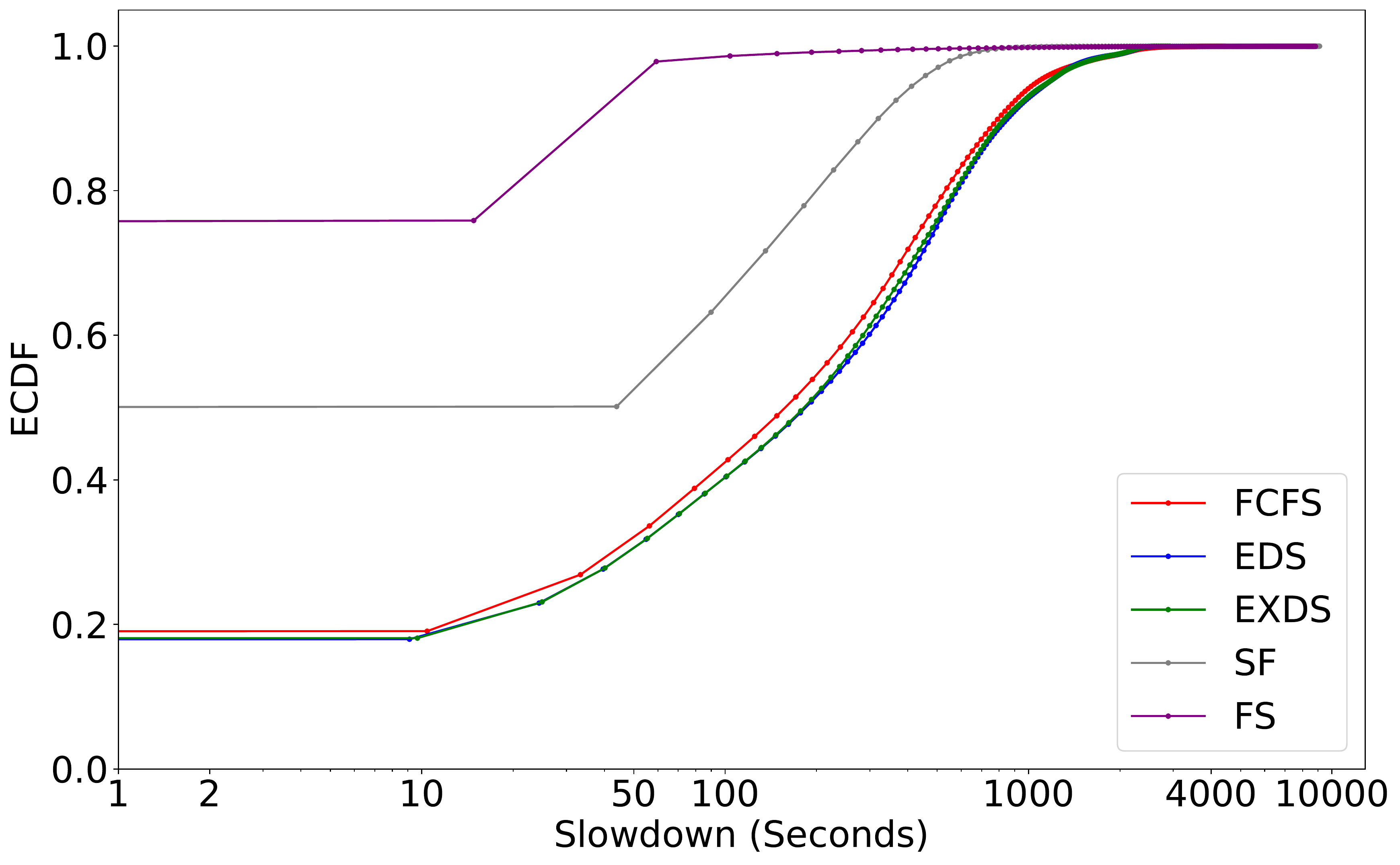}
    \caption{Plot of the slowdown of request execution order policies.}
    \label{fig:queue-policy-plots}
\end{figure}

Figure \ref{fig:queue-policy-plots} displays the outcomes of the request execution order policies as plots. The empirical cumulative distribution function (ECDF) values are on the y-axis and slowdown on the x-axis. X-axis uses log scale. The distribution of the slowdown in FCFS graph resembles the ED graph. ED graph has the least 99th percentile slowdown that is about 2700 seconds, while the FCFS plot reaches a greater value of about 4300 seconds. The SF graph has the highest 99th percentile slowdown value at approximately 9200 seconds, closely followed by the FS graph at about 8800 seconds.

\begin{mdframed}[backgroundcolor=gray!20] 
    \textbf{O-1:} The distribution of the slowdown is similar for first come first serve policy (FCFS) and early deadline policy (ED). However the 99th percentile of FCFS is about 1.4 times the 99th percentile slowdown of ED.
\end{mdframed}

SF has the highest 99th percentile because in the setup most requests have small execution time, since the SF policy gives priority to smaller requests, a few requests with relatively larger execution time are delayed till the end. This results in higher wait time, thus higher slowdown. On the other hand, when using the ED policy requests with a longer execution time will have a larger deadline; thus they will be delayed for some time as smaller requests with shorter deadlines get executed. However, eventually the longer execution time request will get executed as their deadline approach. It is not delayed till the end. Therefore, the slowdown values do not get very high.

\begin{mdframed}[backgroundcolor=gray!20] 
    \textbf{O-2:} Highest 99th percentile slowdown (9200 seconds) for shortest first (SF) policy while least 99th percentile slowdown (2700 seconds) for early deadline policy (ED).
\end{mdframed}

Another way to analyze the plots is to study the distribution of slowdown values. The distribution of FS policy plot is better since around 75\% of requests have minimal slowdown, less than 10 seconds. The next 24\% of requests also experience mild slowdown. Only the last 1 percent of requests, which represent requests with high execution time, have high slowdown. The SF policy is on the second best in comparison as only 50\% of requests have low slowdown. The majority of the numbers in the remaining 50\% range are in the mid-range. However, as compared to FS, the outcome of SF is lacking. The main difference between the FSFC and EDS distributions is the termination point. For FCFS, the slowdown distribution is the worst, with just 20\% of requests experiencing low slowdown while the majority of the remaining requests experience high slowdown.

\begin{mdframed}[backgroundcolor=gray!20] 
    \textbf{O-3:} Fair share (FS) of resources policy performs better compared to other policies as 75\% requests experience low slowdown (less than 10 seconds). For FCFS, the slowdown distribution is
    the worst, with only 20\% of requests experiencing low slowdown.
\end{mdframed}

The two variants of early deadline policy, EDS and EXDS, nearly overlap because the model used for execution time. In the experiment execution time is sampled from the log-normal distribution, which has a heavy tail and less spread. As a result, the majority of requests have close to the same execution times. The same values of execution time at microservice stages results in slack to be equally distributed for the EXDS policy as well. Therefore it results same as the EDS policy. A different experimental setting will be needed to demonstrate the differences between the EDS and EXDS policy.

\begin{mdframed}[backgroundcolor=gray!20] 
    \textbf{O-4:} The two variants of early deadline policy, equal division of slack (EDS) and execution based diviosn of slack (EXDS) almost overlap under given setup. 
\end{mdframed}

\section{Experiment 2: Comparing Load Balancing Policies} \label{section:experiment-2}

The load balancer is used to determine the instance of the microservice to route the request to. A load balancer makes sure no one instance is overloaded. Load balancing increases the responsiveness of an application by distributing the work \cite{load-blancing-benefit-2}. Microservices are also more readily available as a result. There are different load-balancing algorithms. Through this experiment, three load-balancing algorithms are compared which are described in Section \ref{simulator-implementation}. The Section \ref{section:experiment-setup} includes information about the simulator's setup. The difference in setup is that in this experiment the queue policy is fixed to first come first serve and load balancing algorithms are varied. 

\begin{figure}[]
    \centering    \includegraphics[width=.7\textwidth,keepaspectratio]{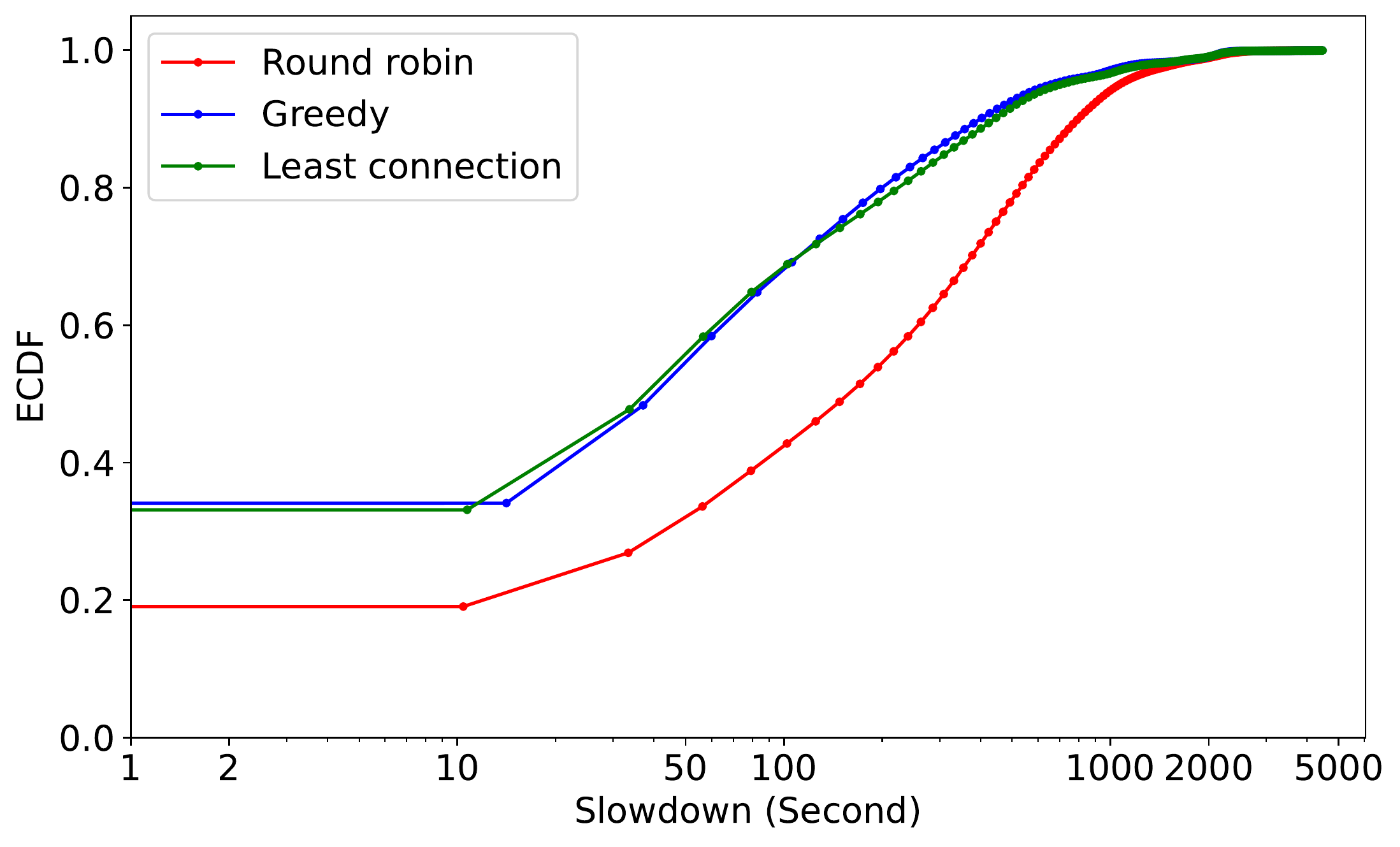}
    \caption{Plot of the slowdown of load balancing policies.}
    \label{fig:load-balancer-results}
\end{figure}

Figure \ref{fig:load-balancer-results} displays the results of the load balancing algorithm as a plot. The log scale of the slowdown in seconds is on the x-axis, while the ECDF values are on the y-axis. Every policy plot ends at about the same time, which is around 4300 seconds. In the current setup, the outcomes of the greedy algorithm and the least connection are identical. The greedy algorithm plot's shape indicates that, while not much better than the least connection policy, greedy performs just marginally better. The greedy algorithm differs in that it needs to know the execution time, which may be derived using a prediction model. While the least connection only requires the quantity of requests that are awaiting, which is easier known than execution time. The other two policies perform better than round robin since they have 35\% of requests with low slowdown values unlike round robin where the percentage is 20.

\begin{mdframed}[backgroundcolor=gray!20] 
    \textbf{O-5:} Similar distribution of slowdown is observed for the least connection load balancer and greedy algorithm. Overall greedy algorithm performs better, followed by the least connection load balancer.
\end{mdframed}

The load balancing policies are further compared using imbalance. In this setup the first microservice has four instances. Focusing on this microservice, we evaluate the imbalance. The imbalance is the mean standard deviation of utilization values of the microservice instances. For each instance of the four instances, the utilization is recorded at 5 min interval. Standard deviation is calculated and mean is taken. The mean is called as imbalance. As there are three policies, the result is three values of imbalance, one for each policy. The values are shown in Table \ref{tab:imbalance-values}. Higher values mean poorer performance as it means utilization was relatively higher. Round Robin load balancer has the highest value while greedy algorithm has lower value comparatively. However, the least connection can also be preferable because in the current setup the ideal execution values are used for the greedy algorithm while the least connection load balancer does not require execution values so there is a trade off between performance and complexity. 

\begin{mdframed}[backgroundcolor=gray!20] 
    \textbf{O-6:} Higher imbalance for round robin load balancer compared to least connection load balancer and greedy algorithms.
\end{mdframed}

\begin{table}[]
\centering
\begin{tabular}{l|l}
\hline
\textbf{Load balancer}    & \textbf{Imbalance} \\ \hline
Round Robin      & 0.53      \\
Least connection & 0.48      \\
Greedy algorithm & 0.45   \\   \hline
\end{tabular}%
\caption{Imbalance of microservice with 4 instances.}
\label{tab:imbalance-values}
\end{table}

\section{Results and Analysis}

Generally, the performance of microservices depends on their setup. The results are specific to the current configuration, so it is not possible to articulate that one policy is superior to another based on the information available. A way to compare the results can be by categorising the policies. The policies analysed in the experiments can be categorised into two categories. The first category is policies that need execution time example greedy algorithm. The second categories are policies that do not require the execution time, example least connection policy. 

The results of the first experiment are, if the goal is to lower the maximum slowdown, there is a trade-off with increasing the overall slowdown. Early deadline policies give better results if the overall increase in the slowdown is within SLA limits. On the other hand, if one percent of requests with high slowdown values can be ignored, then the FS policy is a better policy as it results in reduced slowdown for most of the  requests. For example, the FS policy may be useful for a download service, since high slowdown is reasonable when downloading larger files, while also serve other incoming requests.

The second experiment results show that round robin has the lowest performance among the three policies. The greedy algorithm performs better compared to other load balancing policies as it has the least imbalance. Furthermore, the slowdown distribution shows requests experience lesser slowdown compared to other policies. However the experiment uses ideal execution values to calculate instance load in the greedy algorithm. In production there is always an error in the prediction of execution time, such a greedy algorithm may perform lower. Moreover, we do not take into account the time required for the execution time prediction model to predict execution time for every microservice request. Therefore the preferred choice among the three policies is the least connection under the setup as it only requires the number of waiting requests, which
is easier known than execution time. Additionally, the slowdown distribution of the least connection load balancer is similar to the greedy algorithm. Prior studies have built microservice execution frameworks that use prediction models to predict microservice execution times \cite{prediction}, for example, using machine learning \cite{GrandSLAm}. This approach for selecting microservice instance needs to be discussed further, as our observations imply that the relatively simpler least connection load balancer may be a better option.


}




{\let\cleardoublepage\relax \chapter{Related Work and Limitations}


\ifpdf
    \graphicspath{{6_relatedwork/figures/PNG/}{6_relatedwork/figures/PDF/}{6_relatedwork/figures/}}
\else
    \graphicspath{{6_relatedwork/figures/EPS/}{6_relatedwork/figures/}}
\fi

%

This chapter mentions the related work done by prior studies in the first section. The second section answers the fourth research question about validation of the simulator. 

\section{Related Work}

This section presents the similar work done by prior studies. The related work can be divided into three parts. The first part describes microservice, their motivation and challenges. The paper ``Microservices: yesterday, today and tomorrow'' \cite{ms-arch-review} reviews the history of software architecture, the reasons that led to the diffusion of objects and services first, and microservices later. It provides an evolutionary presentation to help understand the main motivations that lead to the distinguishing characteristics of microservices. 

The second part is the microservice architecture designs in older studies, the architecture designed in the current studies puts together the common components of the architectures from mentioned studies resulting in the microservice architecture in Figure \ref{fig:ref-architecture}. A reference architecture is given in the paper Continuous software engineering—A microservices architecture perspective \cite{continuous-software-engineering}. It includes details on the categories of microservices in the architecture. Another architecture by Microservices Architecture For Modern Digital Platforms \cite{modern-architecture}, it has many detailed components.

Lastly, we discuss the microservice simulators found in previous papers. There are limited studies that build a microservice simulator. One is µqSim \cite{qSim}, a queuing network simulator designed for interactive microservices. µqSim explicitly models each application’s execution stages, and accounting for queuing effects throughout execution, including request batching. The current study simulator is different as it is focusing on simulator design and implementation. On the other hand, µqSim focuses on queuing effects and execution phases.

The current study experiments with request execution order policies. The fourth policy uses slack based division to set the deadline, see Section \ref{section:experiment-1}. The GrandSLAm paper \cite{GrandSLAm} also uses slack based division. However, it focuses more on batching which is beyond the scope of the current study. In this paper we study the slack-based division policy without batching.

\section{Threats to Validity} \label{limitations}

This section describes the validation method used and the limitations of the simulator. Firstly, there is no known validated microservice simulator to compare and validate the current simulator. Another method of validation is to run a real microservice application and compare with simulator results. However such a setup would be complex and goes beyond the scope of the study. In this study mathematical analysis is used to validate the simulator. The result of using the statistical models implemented in the simulator can be mathematically calculated. These results can then be compared with the output of the simulator. For example, for the first-come first serve policy, the average wait time of the requests can be mathematically calculated given the execution time. The calculated value matched with the average wait time output of the simulator. The same method was applied to validate using the shortest job first policy.

Limitations of the simulator include that there is currently no scaling. The results of the experiments would be different with scalable microservice instances. The simulator assumes zero delay at the router which includes the load balancing and service discovery operations. Another assumption is the latency of the request going to, and from microservices is ignored. There is no timeout for requests. Therefore execution time should be carefully set for requests because if somehow a requests execution time is six hours then other requests would wait if using FCFS execution order. This leads to unexpected results. 

The microservice simulator can be used to experiment microservice based application performance under various settings. The current simulator does not use trace as currently microservices trace is not available; instead the simulator uses statistical models to generate workloads. The models can be replaced as we provide a simple interface. Therefore, the simulator can be adapted to use trace as discussed in Section \ref{requirements}.

}





{\let\cleardoublepage\relax \chapter{Conclusion}


\ifpdf
    \graphicspath{{8_conclusion/figures/PNG/}{8_conclusion/figures/PDF/}{8_conclusion/figures/}}
\else
    \graphicspath{{8_conclusion/figures/EPS/}{8_conclusion/figures/}}
\fi

%



This chapter summarizes the work and contributions of the study. Later the answers to the research questions are mentioned. In the end future work options are discussed.

Due to the rising need for scalable applications and distributed software systems, microservice-based architecture have become a blueprint for many large companies and big software systems. At a large scale, interdependencies between microservices can lead to complex microservice systems. Due to this, predicting the performance of microservices can be a challenge. Furthermore, the advantage of microservices over monolith depends on the quality of building microservices. Simulation enables exploration, analysis, and comparison of microservice systems.

We designed a reference architecture for microservice-based applications. Then design decisions were discussed, resulting into a design for a simulator. After which the simulator was implemented in OpenDC. Some appealing features build in the simulator include concurrent execution of microservices, configurable depth, multiple load balancing policies and request execution order policies. The paper includes two experiments that show the simulator in action. The results of experiments show request execution order policy and load balancing policy if properly selected can improve the performance of microservices. Overall the simulator is a great tool that can provide valuable insights about performance of applications using the microservice architecture.

\section{Answering the Research Questions}

The study answers four research questions:

\begin{enumerate}

\item[RQ1] \textbf{What is a good microservices architecture for the simulator? How does it compare to architectures designed in prior studies?}

The requirements for microservices architecture are analysed. The components of the architecture are discussed based on the requirements. Designs from prior studies are studied and evaluated. The microservices architectures designed by prior studies do not meet the requirements. Therefore a \textit{new} microservice architecture is designed to meet the requirements. The architecture contains common components from the prior studies.

\item[RQ2] \textbf{How to implement the components of architecture from RQ1 in the simulator in OpenDC?}

The reference architecture from RQ1 is very abstract. It cannot be directly implemented. Therefore, we build a new simulator design which focused on implementing microservices in the OpenDC, a discrete-event simulator for datcenters. Initially the simulator design requirements are analysed. There are three major features implemented in the simulator. The first feature is that the request can have configurable depth and at any point microservices can communicate to any number of microservices. The second feature is that the simulator will allow comparison of load balancing policies. The third feature is simulator implements four request execution order policies. The simulator allows custom workload with configurable interarrival time, execution time, etc. The details such as the internal components of the simulator and methodology for implementing the components are discussed. The result is a simulator design that can be implemented in OpenDC. The simulator design provides an overview of the simulator. It is then followed to implement the simulator. 

\item[RQ3] \textbf{What is the impact of request reordering and load balancing policies on the performance of microservice applications?}

Focusing on the new, microservice-related features in OpenDC, two experiments are conducted with the simulator. The first experiment compares load balancing policies across microservice instances. The second experiment compares request execution order policies at microservice instance. In both experiments, the project uses the first feature of depth in setup, in particular requests with different depths are used. At the end of the simulation, the simulator outputs a report. The report contains results such as microservice utilization, total time, wait time, execution time, slowdown, etc. The results can be compared to analyse microservices performance. 

Generally, the performance of microservices depends on their setup. The results of the first experiment show if one percent of requests experiencing higher slowdown can be ignored, then the fair share of resources policy (FS) performs better compared to other policies as 75\% requests experience low slowdown (less than 10 seconds). On the other hand, if higher values of slowdown cannot be ignored, then early deadline policy performs better with the least 99th percentile slowdown. The second experiment results show that round robin has the lowest performance among the three policies. The greedy algorithm performs better under ideal conditions. Overall, the preferred choice among the three policies is the least connection load balancer.

\item[RQ4] \textbf{How to validate the simulator to show the results are trustworthy?}

Three methods for validating the simulator are discussed in Section \ref{limitations}. In the study, the validation of the simulator is done using the third approach that is mathematical analysis. The simulator output is matched with mathematically calculated result. 
For example, for the first-come first serve policy, the average wait time of the requests can be mathematically calculated given the execution time. The calculated value matched with the average wait time output of the simulator. The same method was applied to validate using the shortest job first policy. This is minimal validation and has limitations such as only results that have a mathematical formula can be verified. Overall, we find the simulator we designed and implemented does not meet the requirements for complete validation because of the limitations described in Section~\ref{limitations}.

\end{enumerate}

\section{Future Work}

This section describes some ways in which simulator can be used in future studies and for further microservice performance analysis. The microservice simulator can be used for future work because simple replaceable interfaces are provided in the simulator. For example, the log-normal distribution used for execution time of microservices can be replaced with a different model. These models can be compared and analysed.

As mentioned in Section \ref{limitations}, currently no public microservice trace is available. Therefore, the simulator uses statistical models for workload generation. In future, if the trace is available, the simulator can be adapted to use the trace as input for the simulator. The details of microservice trace format are discussed in Section \ref{requirements}.

The simulator can be improved by adding more features. For example, currently the simulator does not support scaling of microservice instances. Features such as auto-scaling can be added for better analyses of microservices performance. Similarly, the simulator simulates asynchronous communication between microservices. It would be interesting to analyse and compare the results when using synchronous communication between microservices. Moreover, cost report for deploying and running microservices is a useful feature that can be added to the simulator.

Further analyses of microservices include comparing microservices with monolith applications. For example, using simulation to analyse performance of the same application as a monolith compared to microservices approach. 

A microservices application consists of tens or even hundreds of services. Each service is a mini-application with its own specific deployment, resource, scaling, and monitoring requirements. What is even more challenging is that despite this complexity, deploying services must be fast, reliable and cost-effective. There are several microservice deployment patterns, including Service Instance per Virtual Machine
and Service Instance per Container \cite{microservices-design-to-deployment}. Another intriguing option for deploying microservices is AWS Lambda, a serverless approach. These microservice deployment strategies can simulated and analysed to further study microservice performance.

}







\bibliographystyle{Latex/Classes/PhDbiblio-url2} 
\renewcommand{\bibname}{References} 




\appendix


\ifpdf
    \graphicspath{{10_appendix/figures/PNG/}{10_appendix/figures/PDF/}{10_appendix/figures/}}
\else
    \graphicspath{{10_appendix/figures/EPS/}{10_appendix/figures/}}
\fi


\chapter{Reproducibility}
\section{Abstract}

This appendix details information on the running the simulator build in this thesis.

\section{Artifact check-list (meta-information)}

{\small
\begin{itemize}
  \item {\bf Program:}
  OpenDC Microservice simulator
  \item {\bf Compilation: }
  Kotlin, JDK. 
  \item {\bf Metrics: } slowdown.
  \item {\bf Output: } microservice performance metrics.
  \item {\bf Experiments: }
  Comparing request execution order policies and load balancing policies.
  
  \item {\bf How much disk space required (approximately)?: }
  Simulation of 8 hr with depth 2, 10 instances in total, runs smooth with heap space of 512 MB. More complex setup would require configuration of more heap space for the simulator.
  
  \item {\bf How much time is needed to prepare workflow (approximately)?: }
  Simply select the distributions the simulator should use.
  
  \item {\bf How much time is needed to complete experiments (approximately)?: }
  Few seconds to minutes depending on time and setup of simulator.
  
  \item {\bf Publicly available?: }
  Yes
  
  \item {\bf Code licenses?: }
  MIT License
  
\end{itemize}
}

\section{Description}

Source code available at \url{https://github.com/magh3/opendc} 

\section{Installation}

Deploy instructions at \url{https://github.com/magh3/opendc#readme}

\section{Evaluation and expected results}

At the end of the simulation results are output. The metrics for router, individual microservice instances and overall simulator metrics are included in the output.

\newpage
\chapter{Self Reflection}

I learned about microservice architecture through this project. I gained knowledge in architectural design. Choosing which component to add, weighing simplicity, complexity, and simulator needs. I learned of the many elements of the microservice architecture. My coding abilities grew as a result of creating and debugging the simulator. I gained knowledge of comparing policies, conducting experiments, and evaluating the outcomes.









\end{document}